\documentclass{emulateapj}
\usepackage{apjfonts}

\usepackage{amsmath,amssymb}
\usepackage{xspace}
\usepackage{graphicx}

\newcommand\LIR{L$_{IR}$}
\newcommand\Lx{$L_{2-10 keV}$}
\shorttitle{Environment of LIRGs in COSMOS}
\shortauthors{Feruglio et al.}

\begin{document}


\title{Obscured star--formation and environment in the COSMOS field}

\author{C. Feruglio\altaffilmark{1}, 
H. Aussel\altaffilmark{1}, 
E. Le Floc'h\altaffilmark{2}, 
O. Ilbert\altaffilmark{3}, 
M. Salvato\altaffilmark{4,5},  
P. Capak\altaffilmark{4}, 
F. Fiore\altaffilmark{6}, 
J. Kartaltepe\altaffilmark{2}, 
D. Sanders\altaffilmark{2}, 
N. Scoville\altaffilmark{4}, 
A. M. Koekemoer\altaffilmark{7} 
Y. Ideue\altaffilmark{8}, 
}

\altaffiltext{1}{DSM/Irfu/Service d'Astrophysique, CEA  Saclay, F 91 191 Gif-sur-Yvette}
\altaffiltext{2}{Institute for Astronomy, 2680 Woodlawn Drive, University of Hawaii, Honolulu, HI 96822.}
\altaffiltext{3}{Observatoriore de Marseille-Provence, Pole de l'Etoile Site de Chateau-Gombert, 38, rue Frederic Joliot-Curie
13388 Marseille cedex 13, France}
\altaffiltext{4}{California Institute of Technology, MS 105-24, 1200 East California Boulevard, Pasadena, CA 91125.}
\altaffiltext{5}{Max planck Institue fur plasma Physics (IPP) and Excellence Cluster, Boltzmannstrasse 2, Garching 85748, Germany}
\altaffiltext{6}{INAF - Osservatorio Astronomico di Roma, via Frascati 33, 00040 Monteporzio Catone, Italy}
\altaffiltext{7}{Space Telescope Science Institute, 3700 San Martin Drive, Baltimore, MD 21218}
\altaffiltext{8}{Ehime University, Japan}
\email{chiara.feruglio@cea.fr}

\begin{abstract}
We investigate the  effects of the environment on star--formation in a sample of massive luminous and ultra--luminous infrared galaxies (LIRGs and ULIRGs) with S(24 \micron)$>80\mu$Jy and $i^+<24$ in the COSMOS field.
We exploit the accurate photometric redshifts in COSMOS to characterize the galaxy environment and study the evolution of the fraction of LIRGs and ULIRGs in different environments in the redshift range $z=0.3\div1.2$ and in bins of stellar mass. 
We find that the environment plays a role in the star formation processes and evolution  of LIRGs and ULIRGs.
We find an overall increase of the ULIRG+LIRG fraction in an optically--selected sample with increasing redshift, as expected from the evolution of the star formation rate density.  
We find that the ULIRG+LIRG fraction decreases with increasing density up to $z\sim 1$, and that the dependence on density flattens with increasing redshift. 
We do not observe the reversal of the star--formation rate density relation up to $z=1$ in massive LIRGs and ULIRGs, suggesting that such reversal might occur at higher redshift in this infrared luminosity range.
\end{abstract}

\keywords{galaxies: general --galaxies: evolution -- large--scale structure of universe -- surveys}

\section{Introduction}
Galaxy properties such as morphology, color, star--formation rate do vary with redshift, stellar mass and environment  (Kauffmann et al. 2004, Postman et al. 2005, Cucciati et al. 2006, Cooper et al. 2008).
In the local Universe,  large redshift surveys (SDSS) have shown that  the star--formation properties of galaxies are dependent on the local environment both in clusters and in low density environments, such as cluster outskirts and the field. 
Red passive early--type galaxies (E+S0) are more frequently found in high density environments, while blue late--type, star--forming  galaxies tend to dominate in low density regions (Dressler 1980). The average star--formation rate (SFR) decreases with increasing density (SFR--density relation), suggesting a scenario where  galaxy interactions, and in general galaxy environment, play a major role in regulating star--formation.
AGN activity is also linked to the environment. In the local Universe luminous AGN are found preferentially in the field and  in underdense regions, where their hosts can preserve large amounts of gas to feed the black--hole (Kauffmann et al. 2004).
Using the SDSS data, Hwang et al. (2010) found no dependence of the fraction of LIRGs plus ULIRGs among infrared galaxies versus galaxy density at fixed stellar mass. Conversely, the star formation activity of LIRGs and ULIRGs are found to strongly depend on the morphology and the distance to the nearest neighbor galaxy, indicating that galaxy-galaxy interactions/merging play a critical role in triggering the star formation activity of local LIRGs and ULIRGs.
The morphology--density and color--density relations are observed at intermediate redshift and apparently are in place by $z\sim 1$ (Cassata et al. 2007, Capak et al. 2007, Tasca et al. 2009, Postman et al. 2005, Cucciati et al. 2006).
Recently,  Scoville et al. (2007b) found correlations between environment and stellar mass, SFR, SED type and morphology  for a large mass--selected sample over the redshift range $z=0.1\div1$ in the COSMOS survey (Scoville et al. 2007a).
Kovac et al. (2009) showed that  star--formation and color transformation rates of galaxies  are higher  in the group environment than outside groups by a factor of 2 to 4 since $z\sim 1$.
Elbaz et al. (2007) and Cooper et al. (2008) observed the reversal of the SFR--density relation at $z\sim 1$: the average SFR increases with increasing density. 
An enhancement of the star--forming fraction at intermediate densities has also been reported for a $z\sim 0.8$ cluster  by Koyama et al. (2008). Tran et al. (2010) found an increasing fraction of IR luminous galaxies with increasing galaxy density in cluster at redshift $\sim$1.6.
Caputi et al. (2009) analyzed the close environment, on 1 Mpc scales, of luminous infrared galaxies (LIRGs, \LIR$=10^{11}\div 10^{12} $L$_{\odot}$) and ultra--luminous infrared galaxies (ULIRGs, \LIR$>10^{12} $L$_{\odot}$)  in zCOSMOS (Lilly et al. 2007),
finding that luminous infrared galaxies  between $z=0.6\div 1$ are more often found in overdense environments, while  ULIRGs prefer underdense environments.
Ideue et al. (2009) found an increasing fraction of [OII] emitters in the high density environments at $z\sim 1.2$ in the COSMOS field, based on narrow band imaging.

All these results suggest that  galaxy evolution and star--formation activity are dependent on the environment, and at z$\gtrsim1$ galaxies are caught in their active star--forming evolutionary phase in high density regions such as the center of groups.
This phase might precede the formation of red ellipticals observed in the cluster centers in the local Universe.   
Both the SFR and the color--density relations exhibit strong stellar mass trends  (Kauffmann et al. 2004,  Balogh et al. 2004, Haines et al. 2007). Therefore, in order to isolate environment from mass--driven evolution (downsizing), it is crucial to study galaxy properties in bins of mass.

Galaxy interactions, fly--bys and minor mergers are among the mechanisms  expected to have an impact on evolution and star--formation activity of galaxies.
Ram pressure stripping by the intra--cluster medium (ICM) might also contribute to regulate star--formation processes in clusters (Kapferer et al. 2008a,b). However observations of intermediate redshift clusters tend to exclude ram pression stripping as a mechanism of star--formation quenching (Demarco et al.).
Disentangling the effect of the cosmological decline of the SFR from environmental effects is not straightforward   (Lilly et al. 1996, Madau et al. 1996, Le Floc'h et al. 2005, Hopkins et al. 2006, Caputi et al. 2007).
Large unbiased samples are needed to isolate the global evolution of the SFR and mass--driven effect (Bell et al. 2004) from environment--driven effects. Deep, large--area surveys, such as COSMOS (Scoville et al. 2007a), are useful to address this topic, as they can sample large volumes of Universe from moderate to high redshifts. 
In particular, the COSMOS large area (2 square degrees in the sky, corresponding to physical scales of  $\sim 40$ Mpc at  $z=1$) allows probing,  from intermediate to high redshift, a wide range of stellar masses and galaxy densities, from the dense clusters and groups to the field. 
This is usually not achieved with deep pencil beam surveys. However, for such large area a complete deep spectroscopic surey cannot be easily achieved. 
For this reason the COSMOS collaboration acquired deep and accurate photometry in (so far) 34 bands from far--UV to mid--infrared, including imaging in 16 intermediate and narrow band filters from the Subaru Telescope (COSMOS--21, Taniguchi et al. 2007, Capak et al. 2009).  
The survey has, on average, a photometric point every 200 \AA~  between the U and z bands (29 photometric bands in this wavelength range), simulating low resolution spectroscopy. Thus, highly reliable photometric redshifts are easily computed (Ilbert et al. 2009a, Salvato et al. 2009).  As demonstrated by Scoville et al. (2007b), at redshift below 1.2 the quality of the photometric redshifts is sufficient to achieve the accuracy needed for this work, i.e. tracing the evolution of galaxies, AGN and star--formation processes in the large scale structure.

In this paper we use the COSMOS dataset to characterizing the environment of bright star--forming galaxies , and studying their properties as a function of the environment. 
The plan of the paper is as follows:  Sect. 2 presents the multiwavelength dataset. Sect. 3 presents the sample selection, while Sect. 4 presents the method for density computation. Sect. 5 presents the galaxy properties as a function of the environment, and Sect. 6 presents a summary of the results. 
A standard cosmology ($\Omega_m = 0.3$ and $\Omega _{\lambda}=0.7$) with $H_0 = 70$ km s$^{-1}$ Mpc$^{-1}$ is used throughout. Magnitudes are given in the AB system.

\begin{figure}
\includegraphics[scale=.4]{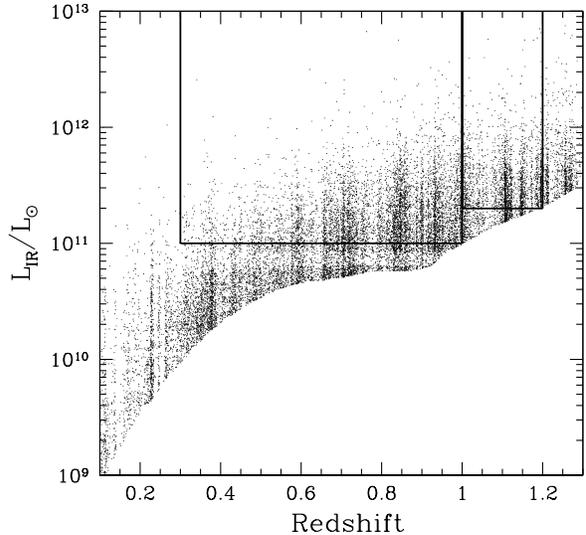}
\caption{$L_{IR}$ versus redshift  for the 24 \micron\ sample with S(24 \micron\  )$>80~\mu$Jy. Solid lines show the luminosity  and redshift selections.}\label{zlir}
\end{figure}

\section{Data}
The COSMOS field has deep multi--wavelength photometric coverage over an area of 2 square degrees from X--rays to the radio. 
The photometric catalog includes data from 34 broad, intermediate and narrow bands from 1550 \AA\ to 8 \micron\ ,and it is complete down to $i^+ =26.2$ (5$\sigma$ for point a point source, Capak et al. 2009).
We use the  photometric redshifts drawn from the catalogs of  Ilbert et al. (2009a).
These have an accuracy as good as $\sigma_z/(1+z)=0.012$ up to $i^+=$24 and z$\leq 1.25$.  For fainter $i^+$ magnitudes the accuracy of photometric redshifts rapidly deteriorates  (Ilbert et al. 2009a).
This paper is limited to galaxies with $i^+<24$, i.e. the sample with the best photo--z accuracy available.
Ilbert et al. (2009a) photo--z were complemented by those computed by  Salvato et al. (2009) for the AGN detected in the XMM--Newton survey of COSMOS (Cappelluti et al. 2009, Brusa et al. 2007).
The latter account for AGN time--variability and use a library of composite template SEDs, including both AGN and galaxy contributions. 
The precision of AGN photo--z is $\sigma_z/(1+z)=0.014$ for $i^+<24$ at any redshift.

The stellar masses were derived from K--band luminosity and mass--to--light ratio using the Arnouts et al. (2007) relation and Salpeter IMF  (Ilbert et al. 2009b). The measured mass--to--light ratios are consistent within 0.1 dex with the analytical relation of  Arnouts et al. (2007)  for stellar masses larger than $3\times 10^{10} $M$_{\odot}$ for both quiescent and star--forming galaxies (Ilbert et al. 2009b).

\subsection{MIPS 24 \micron\  data}
The full 2 deg$^2$ COSMOS field was observed by Spitzer/MIPS down to flux limits of 80 $\mu$Jy at 24 \micron\ (S--COSMOS,  Sanders et al. 2007).
The multiband catalog of the counterparts of the MIPS 24 \micron\ sources is described in Le Floc'h et al. (2009).
Here we briefly recall its main features.  
The counterparts of the 24 \micron\  sources were first identified by cross--matching with the $K_s$--band COSMOS catalog (McCracken et al. 2010) using a matching radius of 2\arcsec.
The identification was based on the closest $K_s$--band source to the MIPS position, and using the IRAC 3.6 \micron\ as a prior in case of multiple possible  counterparts in the $K_s$--band.
About 90\% of the MIPS sources above 80 $\mu$Jy have been assigned a secure near--IR counterpart, and about 5.5$\%$ have less secure identifications. The final catalog consist of 39360 objects.
The $K_s$--band sources were merged with the photometric redshift catalog of Ilbert et al. (2009a) and Salvato et al. (2009) with a matching radius of 1\arcsec, in order to associate a photometric redshift to each 24 \micron\ source.
Here again the closest $i^+$--band source was chosen as the counterpart of the $K_s$--band source, yielding 90\% of optical identification for the full 24 \micron\ catalog down to $i^+ =26.5$ (see Le Floc'h et al. (2009) for a detailed discussion).
The MIPS 24 \micron\  sample used in this paper includes  AGN detected by XMM--Newton (about 1200 with \Lx$>10^{42} $ erg/s, Cappelluti et al. 2009, Brusa et al. 2007).

Overall, the photometric redshift accuracy for the 24 \micron\ sample is similar to that of the optical sample,  $\sigma_z/(1+z)=0.010$ for $i^+<24$ (Ilbert et al. 2009a). In particular the fraction of catastrophic outliers is $<1$\%  for sources with flux $S(24\micron)>80~\mu$Jy.

\subsection{Star-formation rates}
Chary \& Elbaz (2001) have shown that the mid--infrared flux is a good estimator for the total infrared luminosity (L$_{IR}$) emitted by a galaxy in the range $8 \div 1000$ \micron\ in the local universe.
Our sample is limited to $z < 1.2$, for which the mid--infrared and radio estimators provide consistent L$_{IR}$ values at least in the LIRG regime (Elbaz et al. 2002). 
This assumption is based on infrared spectral energy distributions (SED) typical of luminous star--forming galaxies, whose bolometric output is dominated by star--formation. SFR of sources detected in X--rays may be overestimated due to the AGN contribution to their infrared SED. 

Based on this assumption, we use the mid--infrared flux at 24 \micron\ and Chary \& Elbaz (2001) templates to compute the \LIR.
This method uses a library of 105 templates and the 24 \micron\  flux density to compute the mid--infrared rest--frame luminosity, and derive \LIR\ for each galaxy. 
Figure \ref{zlir} shows the \LIR\ as a function of the redshift for the MIPS 24 \micron\ catalog down to 80 $\mu$Jy.
We convert $L_{IR}$ into the total star-formation rate ($SFR_{IR}$) by using the Kennicutt (1998) relation: 
$SFR_{IR} = 1.72\times10^{-10} $L$_{IR}$[L$_{\odot}]$. 
We neglect the UV contribution since this makes up a small fraction (about 10\%) of the bolometric luminosity for LIRGs at $z\sim 1$ (Burgarella et al. 2007). The SFR are derived assuming a Salpeter IMF, consistent with that used to derive stellar masses.

\begin{table*}
\begin{center}
\caption{Stellar mass completeness.\label{tbl-2}}
\begin{tabular}{crrrrrrrrrrr}
\tableline\tableline
Redshift &$0.3\div0.6$ & $0.6\div0.8$ & $0.8\div1$ &$ 1\div1.2$\\
\tableline
F$_{opt}(>10^{10}$M$_{\odot}$)      &      0.06 (94)                &       0.13  (87)                     &  0.34  (66)  &   0.57 (43)\\
F$_{opt}(>3\times10^{10} $M$_{\odot}$)     &    0.006  (99.4)           &     0.03  (97)                      & 0.22  (78)    &  0.41 (59)\\
F$_{IR}(>3\times10^{10} $M$_{\odot}$)    & 0.001 (99.9)&   0.01 (99)      &  0.06 (94)  &  0.25 (75)\\
\tableline
\end{tabular}
%\enddata
\tablecomments{Fraction of galaxies in the optical and 24 \micron\ samples with $i^+>24$ and stellar mass larger than $10^{10}$ and $3\times 10^{10} $M$_{\odot}$ from redshift $z=0.3$ to 1.2. The corresponding value of stellar mass completeness for $i^+<24$ is given in parenthesis for each bin.}
\end{center}
\end{table*}

\begin{figure}  
\begin{center}
\includegraphics[scale=0.4]{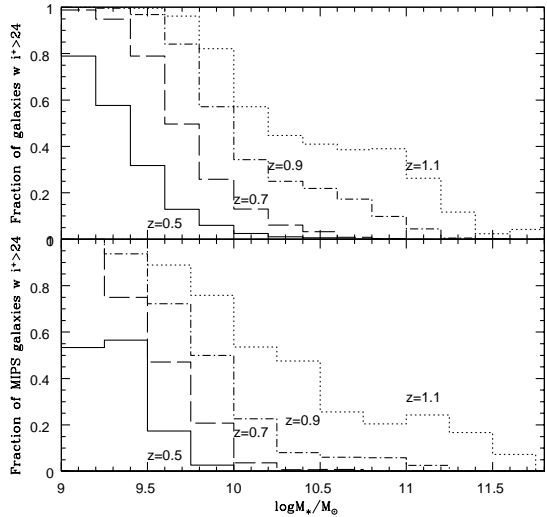}
\caption{Upper panel: fraction of optically-selected galaxies with magnitude $i^+>24$ as a function of the stellar mass and redshift. Redshift is increasing from $z=0.4$ to 1.2, from left to right. Lower panel: fraction of 24 \micron selected galaxies with S(24\micron$)>80~\mu$Jy and  $i^+>24$ as a function of the stellar mass and redshift.}
\label{zlogm}
\end{center}
\end{figure}

\begin{figure}
\includegraphics[scale=.4]{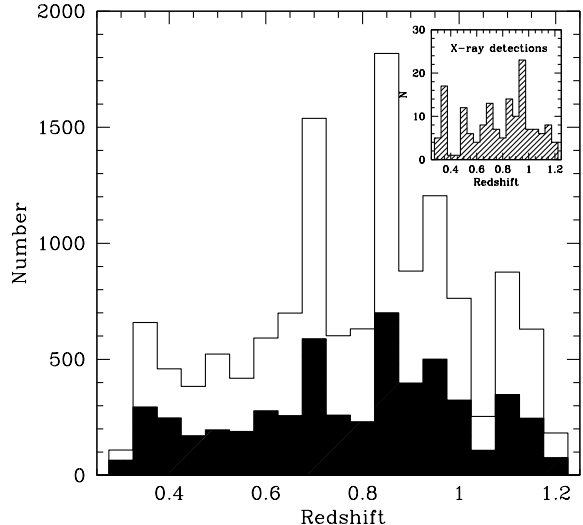}
\caption{Open histogram: all sources with $i^+<$24, $M_{\ast}>10^{10.5} $M$_{\odot}$; filled histogram: LIRGs with the same selection; small panel: LIRGs detected in X--rays by XMM--Newton.  }\label{istoz}
\end{figure}

\section{Sample selection}
The following analysis is based on a mass--selected samples drawn from the MIPS 24 \micron\ and optical catalogues (Sanders et al. 2007, Le Floc'h et al. 2009, Ilbert et al. 2009a) in the $z=0.3\div 1.2$ redshift range.
In order to  assess the completeness of our samples in stellar mass, we compute the fraction of  optically and 24 \micron\ selected galaxies with apparent magnitude $i^+$ larger than a given value as a function of their stellar mass.
Figure \ref{zlogm}, upper panel, shows the fraction of optically selected galaxies with magnitude $i^+>24$ as a function of the stellar mass and redshift, in four redshift bins from $z=0.3$ to 1.2. 
For M$^\ast>3\times 10^{10} $M$_{\odot}$ the fraction of  galaxies with  $i^+>24$ is $<0.01, 0.03, 0.22$ and 0.4 at  $z=0.3\div 0.6, 0.6\div 0.8$, $0.80\div 1$ and $1\div 1.2$, respectively (Table 1).
Therefore for  $i^+<24$ the stellar mass completeness at M$^\ast=3\times 10^{10} $M$_{\odot}$ is  99, 97, 78 and 60\%  in each of the previous redshift bins, respectively. 
The 24 \micron\ detected galaxies are a subsample of the previous one, since they were selected to have an optical counterpart brighter than  $i^+=24$. 
Figure \ref{zlogm}, lower panel, shows the same fraction as before for galaxies with S(24\micron\ )$>80~ \mu$Jy and magnitude $i^+>24$.
The fraction of such galaxies with M$^\ast>3\times 10^{10} $M$_{\odot}$ is $<0.001, <0.1, 0.06$ and 0.25 at $z=0.5, 0.7, 0.9$ and 1.1, respectively. 
Thus, the completeness in stellar mass  is larger than $90\%$ up to $z=1$ and  75\% up to $z= 1.2$. 

The cut in stellar mass adopted here implies that all star--forming galaxies in the sample are LIRGs or ULIRGs, given the relation between SFR and stellar mass at $z\sim 1$ (Elbaz et al. 2007).
We investigate the effects of environment on dusty, star--forming galaxies with $i^+\leq 24$, \LIR$> 10^{11} $L$_{\odot}$ and stellar mass M$^\ast>3\times 10^{10} $M$_{\odot}$ up to $z=1$.
In the redshift bin $z=1\div 1.2$ we will consider galaxies with \LIR$>2\times 10^{11} $L$_{\odot}$ (Figure \ref{zlir}).  %and $M_{\ast}>10^{10.8} M_{\odot}$
This  selection implies  a SFR sensitivity SFR$>18 $M$_{\odot}$/yr and $35 $M$_{\odot}$/yr up to $z=1.2$.
The selected sources are LIRGs and ULIRGs. ULIRGs are a small fraction of the whole sample ($\sim 1.8\%$), therefore we will refer to the sample as LIRGs throughout the paper, keeping in mind that it includes also a small fraction of ULIRGs.
The properties of LIRGs will be compared to those of optically selected galaxies in the same range of stellar mass and magnitude (M$^\ast>3\times 10^{10} $M$_{\odot}$,  $i^+<24$). 
By adopting this selection we focus on the high--mass end of the galaxy mass function (Ilbert et al. 2009b).
The final samples consist of  28455 optically selected galaxies  and 8234 LIRGs in the given redshift range.

The redshift distribution of the two samples is shown in Figure \ref{istoz}, where the X--ray detected AGN are also highlighted.
AGN  with L(2-10 keV)$>10^{42}$ erg/s among the 24 \micron\  detections down to 80 $\mu$Jy and $i^+<24$ represent a small fraction of the population: about 3\% in the redshift range  $z=0.3\div 1.2$  and 4\% at $z=0.8\div 1.2$.
For these the SFR might be overestimated due to the AGN contribution to the infrared SED.
We will show that the trends are not modified if they are excluded in the LIRG sample.
We do not attempt to account for the contamination by obscured, Compton--thick AGN at $z\sim 1$. 
However, we notice that they can hardly outnumber X--ray detected AGN (Fiore et al. 2008, Martinez-Sansigre et al. 2007).  The fraction of highly obscured, not X--ray detected AGN at $z=0.7\div 1.2$ and with intrinsic luminosity $L(2-10 keV)=10^{43.5\div44}$ erg/s is $3.7\times 10^{5} $Mpc$^{-3}$, corresponding to ~67\% of that of X--ray selected AGN (Fiore et al. 2009).

\section{Projected densities}\label{dens}

We adopted the Nth nearest neighbor density estimate (Dressler 1980, Capak et al. 2007, Guzzo et al. 2007,  Postman et al. 2005, Smith et al. 2005).
Projected density measurements based on photometric redshifts are able to detect the structures and reconstruct the overdensities (Scoville et al. 2007b).
They have been successfully used to detect the large scale structure and nicely match the structures detected by X--ray and spectroscopic surveys.
Although limited by the photo--z precision (which implies a smoothing effect), the photometric redshifts trace the dependence of galaxy properties on environment in the LSS and can use large unbiased galaxy samples, a factor of $\sim 10$ larger samples than the largest spectroscopic surveys thus limiting the statistical scatter (Scoville et al. 2007b, Scoville et al. 2010). 
In particular, Scoville et al. (2007b) showed that the correlations between galaxy properties, such as SFR, stellar mass, and environment are not smoothed out when using accurate photometric redshifts. 
Conversely, spectroscopic surveys allow for precise distance measurements but are limited by the sampling rate which is usually low. 
%we use about 10 times more galaxies than spectroscopic samples so reducing the statistical scatter.

The projected density around each galaxy is derived from the distance to the Nth neighbor, {\emph r}, which defines a circular area whose surface density is  $\Sigma_N = \frac{N}{\pi r^2}$.
This density is then divided by the median of projected densities of each redshift slice across the whole COSMOS field, in order to obtain a relative density $\Sigma_N/\overline \Sigma$ in each redshift interval. 
%A background density, computed as the median of projected densities at each galaxy position and redshift slice across the whole COSMOS field,  is subtracted in order to account for  fore/background interloper  contamination due to the uncertainties in the distances. 
We compute projected densities for each galaxy with magnitude $i^+<24$ and stellar mass larger than 10$^{10} $M$_{\odot}$, outside masked areas contaminated by saturated stars (Ilbert et al. 2009a), using 50390 galaxies in total.  
We use a cylinder with a depth of $\pm3\sigma_z$ centered on each galaxy. 
The main difference with previous studies on the COSMOS field is that the accuracy of the photometric redshifts has now improved by roughly a factor of 3. 
The density field of the COSMOS data is measured using the same selection described above.
%optically selected galaxies with $i^+<24$ and  a mass cut--off of M$^\ast=10^{10} $M$_{\odot}$ (see Sect. 3). 
At these stellar masses the completeness is 98, 88, 66 and 43\%  from $z=$0.3 to 1.2 (Figure \ref{zlogm}).
As discussed also in Scoville et al. (2007b), a higher stellar mass cut--off would be more conservative and yield larger completeness, but at the same time would limit the sample to only the most massive galaxies and under--represent lower mass galaxies, therefore limiting  the dynamic range. Therefore when computing densities we allow for a lower completeness. Conversely, when comparing LIRGs to optical galaxies we adopt a cut at $M^\ast>3\times 10^{10} $M$_{\odot}$ in order to have a higher completeness level (see Sect. 3). 

\begin{figure}
%\plottwo{tredg_z09.eps}{testerr.eps}
\includegraphics[scale=.4,angle=0]{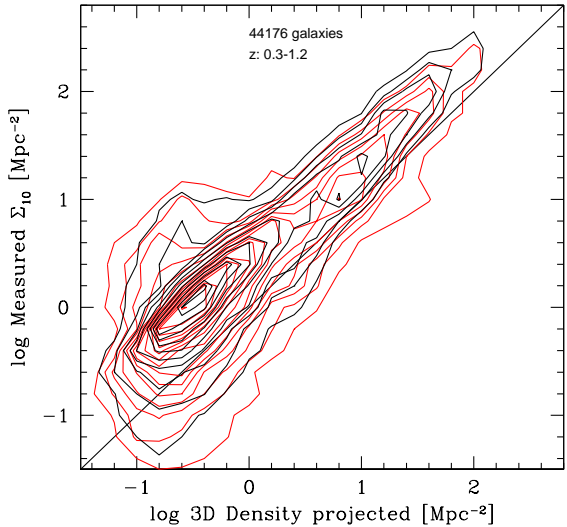}
\caption{Isodensity contours of measured $\Sigma_{10}$ versus projected 3D galaxy density, from the Millennium Simulation (black contours).  Red contours: same relation where the error on photo--z is taken into account.}\label{app} 
\end{figure}

We used the Millennium Simulation mock catalogs (Kitzbichler \& White 2007) to compare the derived surface densities with the three dimensional (3D) volume densities for galaxies selected in the same way as for the COSMOS data.
We run the simulation here described using the same selection used for the observed data.
First, we compute the 3D volume densities by using the positional information available from the Millennium Simulation, and measure the density field around each galaxy in a sphere of radius equal to the distance of the 10th nearest neighbor.
In the simplest assumption, i.e. spherical symmetry, the 3D densities are related to the projected densities by  
 $\rho= 10 \times 3 / 4  \pi r^3_{10}$,
%\delta $ = 10 \times 3 / 4$\pi$ r^3_{10}}$, 
where $r_{10}$ is the distance to the 10th nearest neighbor. 
No redshift errors were included in the 3D density computation at this stage.
This density is  related to the projected density by the following relation: $\Sigma_{10} = \alpha \rho^{2/3} $ where $\alpha = 1.76$.
Figure \ref{app}  shows the correlation between the input and recovered densities. %for all but the sparsest regions
The horizontal axis shows the projected 3D volume density in the Millennium simulation computed as described above. The vertical axis shows the $\Sigma_{10}$ measured in the Millennium simulation in the same way as for the COSMOS data as described at the beginning of this section. 
This test indicates that projected densities on average overestimate the real densities by a factor of 2.2 but there is proportionality between the 2D and the 3D density estimates for the simulated data (Figure 4, black contours). 
In order to estimate how the photo-z errors affect the 2D density computation
we  associate to each redshift in the mock catalog a random error drawn from a gaussian distribution  with $\sigma_z/(1+z)=0.012 $, equal to the error in the measured photometric redshift of COSMOS, and we measure the projected densities.  
A gaussian error distribution is an approximation of the real error distribution.  
However, it is reasonable, considering that catastrophic errors in the photo--z are not an issue both for the optically selected and the 24 \micron\ selected samples down to $i^+=24$, since they are of the order 1\% and will therefore only affect densities by that amount (Ilbert et al. 2009a).
The effect of the photometric redshift error is shown in Figure \ref{app} with red contours. 
The errors in the photo--z increase the scatter in the correlation between 3D and projected densities. However projected densities are still proportional to 3D densities and environment can be characterized at least in the high density regime.
The scatter of the correlation is large at small densities but it decreases with increasing density. This can affect and smooth out possible existing trends by scattering galaxies between contiguous density bins.
At $\Sigma_{10}\sim 10$ the scatter is about 100\% of the density value, meaning that a galaxy residing in a region of density 10$/$Mpc$^2$ will be assigned to a density range $0\div 20/$Mpc$^2$ with a probability of 68.3\%. At $\Sigma_{100}>$100 the scatter is about 25\% of the density value. Therefore, for example, a galaxy residing in a region of 200/Mpc$^2$, has 68.3\% probability of being assigned to a density in the range $150\div 250/$Mpc$^2$. Around 50/Mpc$^2$ the scatter is $\sim 50$\%. This must be taken into account  when studying trends as a function of the environment, and the density bins must be chosen accordingly. 
In particular, for the four density bins used in this paper (see section 5), the 1$\sigma$ error on the density estimate is 150\%, 68\%, 42\% and 32\% of the projected density value,  for each bin from the low density to the high density one, respectively.

\begin{figure*}[t]
\begin{center}
\includegraphics[scale=.5,angle=-90]{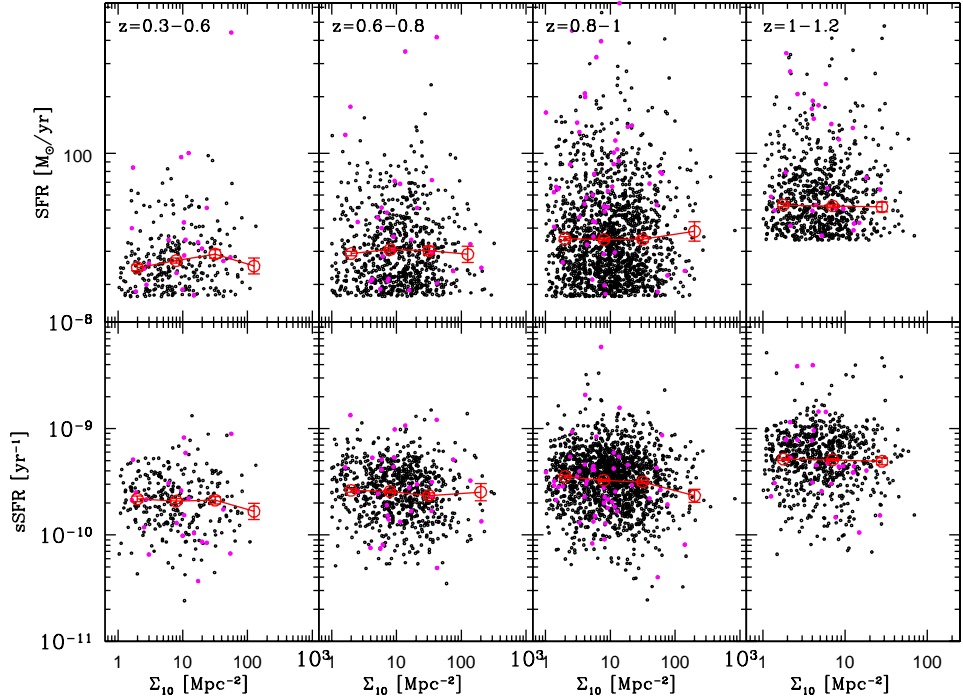}
\caption{ Upper panels: SFR versus density for the LIRG sample (black symbols) and  AGN (magenta symbols) in redshift bins. Red symbols represent the median values and standard errors in density bins, omitting AGN. Lower panels: SSFR versus density.  }\label{figdens1}
\end{center}
\end{figure*}

\section{Results}

\subsection{SFR and SSFR versus environment and redshift}

Figure \ref{figdens1}, upper panels,  shows the SFR of the LIRGs versus the projected density in four redshift bins from $z=$0.3 to 1.2. 
The increase in the median SFR at $z=1\div 1.2$ is due to the luminosity cut adopted here (\LIR $>3\times 10^{11} $L$_{\odot}$).
The increase of the specific star--formation rate (SSFR) with redshift is expected given the evolution of the stellar mass--SSFR relation with redshift (Elbaz et al. 2007).

The SFR--density relation shows a large scatter and the median SFR is constant at all densities.
The median SFR per LIRGs does not vary with the environment. 
In our analysis we do not find any excess of LIRGs in the high density environments up to $z=$1. 
We notice that possible existing trends are likely to be erased in the low density regime (first two bins in projected densities), but the flat trend observed here is significant in the two highest density bins. 
This said, we do not observe the reversal of the SFR--density relation  (i.e. average SFR increasing with density, Elbaz et al. 2007, Cooper et al. 2008) up to $z=1$. 
At $z>1$ the median SFR is compatible with a flat dependence on density.
This difference may be due to the highest accuracy in the density measurements achieved by using spectroscopic redshifts.  
However, the samples of Elbaz et al. (2007) and Cooper et al. (2008) are not mass--limited and spanned a broader range both in stellar mass and SFR, probing down to significantly lower SFR compared to our LIRG sample.
Therefore, such reversal by might be due to the contribution of galaxies with lower stellar mass and SFR. 
Indeed, in the local Universe, the dependence of SFR on environment is weaker for massive galaxies (Kauffmann et al. 2004). 
For massive LIRGs such reversal, if any, might occur at $z>1$.
Figure \ref{figdens1}, lower panels,  shows the SSFR versus density for each of the previous redshift bins.
Here again the SSFR is consistent with a constant trend at all densities.

\begin{figure*}[t]
\begin{center}
\includegraphics[angle=0,scale=0.8]{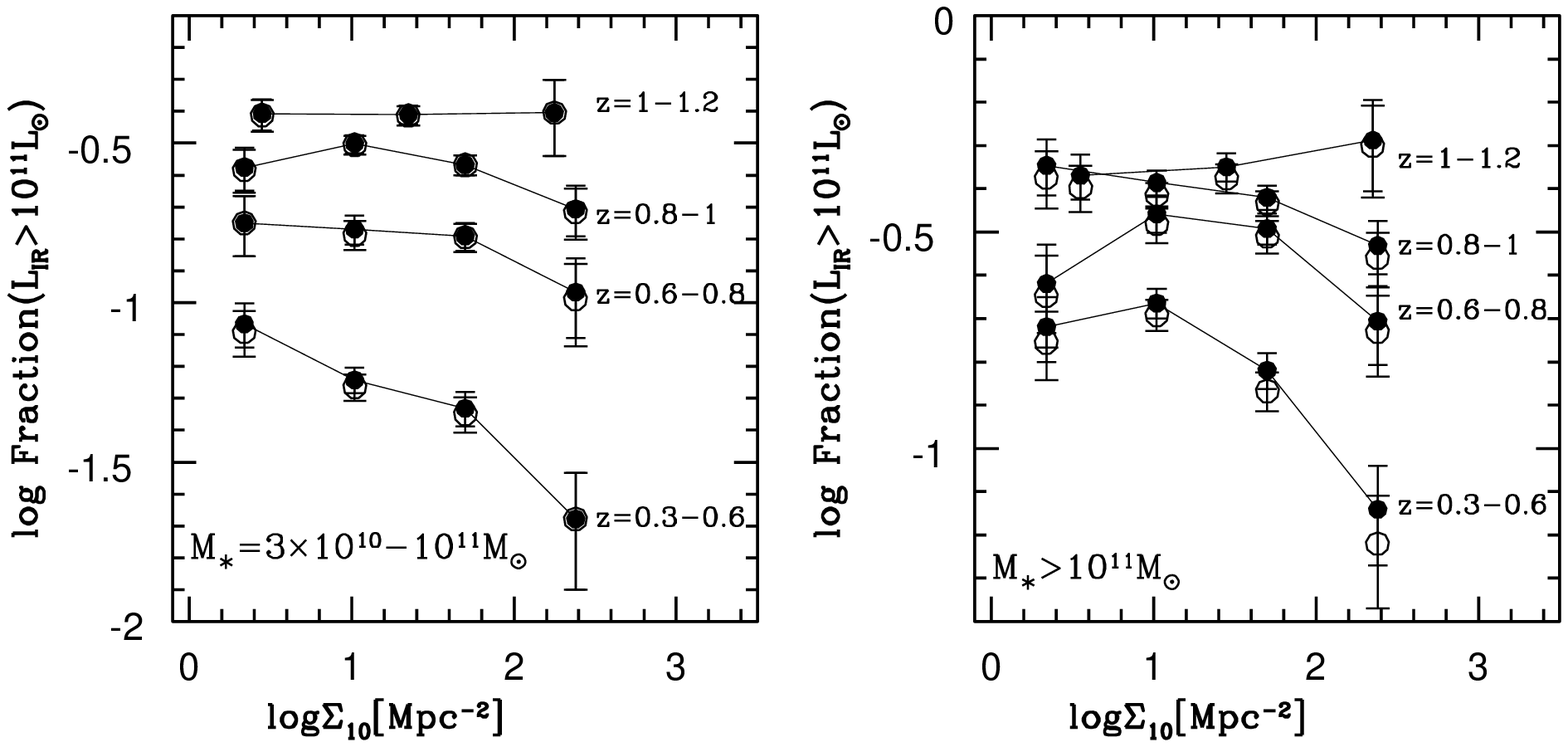}
\caption{ Fraction of LIRGs to the optically selected galaxy population with M$_\ast>3\times 10^{10}$M$_{\odot}$ as a function of  galaxy density and redshift in two stellar mass bins: M$_\ast=3\times 10^{10}\div  10^{11}$M$_{\odot}$ (left panel), M$_\ast> 10^{11}$M$_{\odot}$ (right panel). Filled circles: X--ray detected AGN included, open circles : X--ray detected AGN excluded.}
\label{frlirglowz}
\end{center}
\end{figure*}

%\begin{figure*}[t]
%\begin{center}
%\includegraphics[angle=-90,scale=0.5]{lfrlirg_highz_v3.eps}
%\caption{ Logarithmic fraction of LIRGs with log\LIR$>11.3$ with respect to the whole galaxy population with \logM$>10.8$ as a function of galaxy density and redshift for two stellar mass bins. Symbols as in Figure \ref{frlirglowz}. }
%\label{frlirgh}
%\end{center}
%\end{figure*}

\subsection{LIRG fraction versus environment and redshift}

Figure \ref{frlirglowz} shows the logarithmic fraction of LIRGs among the whole population of galaxies with M$^\ast>3\times 10^{10} $M$_{\odot}$, as a function of the projected density  from redshift $z=0.3$ up to $z=1.2$, and in two stellar mass bins (M$^\ast=3\times 10^{10}-10^{11} $M$_{\odot}$  and M$^\ast>10^{11} $M$_{\odot}$).
For galaxies at $z>1$ the density bins are larger than at lower redshift in order to contain a sample large enough.
Filled circles represent the LIRG fraction among the whole optically selected population with M$^\ast>3\times 10^{10} $M$_{\odot}$. 
Open circles represent the same fraction but where the contribution of X--ray detected AGN has been subtracted from the LIRG sample.
The error--bars on the measurements were computed propagating the poisson errors on the galaxy counts and using upper limits in the bins where we have only a few counts (Gehrels 1986). 
The error--bars are large in the highest density bins, since these are in general less populated (hosting a few tens optical galaxies, and a few LIRGs).

We find that the LIRG fraction correlates with both redshift and environment.
{\emph i}) LIRGs show an overall increase of their fraction with increasing redshift.  Such rise is predicted by the evolution of the LIRG luminosity function(Le Floc'h et al. 2005, Caputi et al. 2007, Magnelli et al. 2009) and of the mass function by spectral type (Ilbert et al. 2009b). 
{\emph {ii}})  We find that the LIRG fraction decreases in the high density environments.
At densities larger than about $80\div 100$ Mpc$^{-2}$ the LIRG fraction drops compared to the low density environment.
The decrease of the LIRG fraction with increasing density is observed up to $z\sim 1.2$, but the dependence flattens with increasing redshift. 
The difference between  the 3rd and the 4th density bin decreases from the low redshift sample (2.0$\pm$0.2 factor at $z=0.3\div 0.6$) to the high redshift ones (1.6$\pm$0.16 and 1.3$\pm$0.1 factors at $z=0.7$ and 0.9, respectively), for stellar masses M$^\ast=3\times 10^{10}-10^{11} $M$_{\odot}$.  
The first two density bins should not be considered since in this range the uncertainties in the density measurement probably erase possible existing trends. 

Figure  \ref{frlirglowz}, right panel, the LIRG fraction is shown for galaxies with stellar masses larger than M$^\ast=10^{11} $M$_{\odot}$.
Again, the LIRG fraction decreases towards high densities, and the decrease appears to flatten with increasing redshift.
The ratio between the third and fourth density bin is 2.15$\pm$0.2 at $z=0.3\div 0.6$ and it decreases to $1.6\pm0.15$ and $1.3\pm0.1$ at $z=0.7$ and 0.9, respectively.
At $z>1$ the LIRG fraction is constant at all densities within the errors for both stellar mass bins.

We note that the dependence is observed also for the LIRG sample cleaned from contribution by the AGN detected by XMM--Newton survey.  
Indeed, the two samples, with and without AGN,  follow the same trend within the errors (Figure \ref{frlirglowz}). 
The environmental properties of AGN in the large--scale structure and the comparison with star--forming galaxies are beyond the aim of this paper, and they  will be addressed in a separate work.

\begin{figure}
\begin{center}
%\includegraphics[scale=0.35]{frmipsmass.eps}
%\caption{ Logarithmic fraction of LIRGs versus the stellar mass in redshift bins. Filled symbols: high density environment ($>$80/Mpc$^{2}$). Open symbols: low density environment  ($<$80/Mpc$^{2}$).}
\includegraphics[scale=0.4]{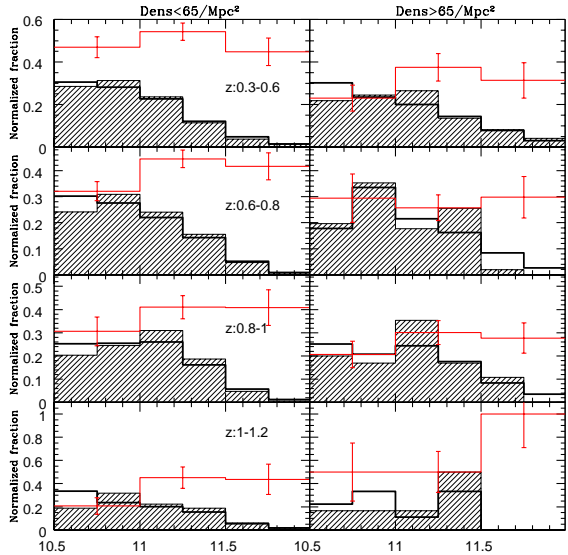}
\caption{Normalized distributions of the stellar mass for the optically selected sample (thick solid histogram) and for LIRGs (shaded histogram), in redshift bins. Left panels show the distributions for projected density $<90/$Mpc$^2$, right panels those for density $>90/$Mpc$^2$. Red histograms show the LIRG fraction to optically selected galaxies, and poisson errors.}
\label{frmass}
\end{center}
\end{figure}

The decreasing fraction observed here is not due to projection effects since: a) the size of the two highest density bins is larger than the average scatter of the relation shown in Figure 4; b) the errors on projected densities tend to smooth out trends by scattering galaxies between contiguous bins. 
We explore the possibility that the decreasing fraction of LIRG in the high density environment might be produced by mass segregation, i.e.  the most massive galaxies are formed in the most massive halos at earlier times, and therefore, preferentially reside in high density environments.
As mentioned before, we studied the environmental dependence in two stellar mass bins in order to minimize such effect. 
In Figure \ref{frmass} we plot the normalized distributions of stellar masses of LIRGs and of the optical sample in redshift bins and in two regimes, high and low projected density ($>$ and $< 90/$Mpc$^{2}$ respectively).  The red histograms show the LIRG fraction, with poisson errors taken into account. Such fraction is plotted in 0.5 dex wide bins of stellar mass in order to minimize the noise.
The mass distribution of LIRGs is not skewed towards lower masses compared to optically--selected galaxies. The Kolmogorov--Smirnov (KS) test shows that the distributions of LIRGs and optical galaxies derive from the same parent population with a probability larger than $98\%$ for all the bins except the last three ($z=1\div 1.2$ and $z=0.8\div 1$, high density), for which the KS test gives no significant result. 
However, the LIRG fraction does not show any significant trend as a function of stellar mass in any environment in the stellar mass range studied here.
Therefore the decreasing fraction of LIRGs in high densities (Figure \ref{frlirglowz})  is not a mass related effect but it is due to the environment.

\begin{figure}
\begin{center}
\includegraphics[scale=0.4]{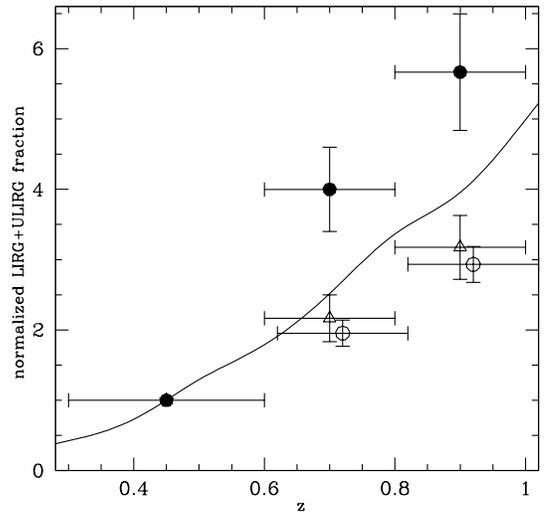}
\caption{Evolution of the normalized LIRG fraction with redshift in 3 redshift bins up to $z=$1. Open triangles: all LIRGs, filled circles: LIRGs in high density environment ($\Sigma_{10}>90$), open circles: LIRGs in low density environment ($\Sigma_{10}=5\div 10$, slightly shifted for clarity), solid line: expected evolution from Magnelli et al. (2009) LF and Ilbert et al. (2009b) MF.  The fractions are normalized to the value at $z=0.3\div 0.6$.}
\label{frmipsz}
\end{center}
\end{figure}

Figure~\ref{frmipsz} shows the fraction of LIRGs as a function of redshift, normalized to their values in our first redshift bin ($z=0.5$).  Empty triangles represent the whole sample, regardless of the environment. The LIRG fraction increases with redshift, reflecting the evolution of the whole infrared selected population. This increase is similar to the global increase observed in Figure~\ref{frlirglowz}. 
We compare it with the expectations from the known evolution of the galaxy population: we use the mass functions of Ilbert et al. (2009b) to compute the expected number of galaxies with M$_{\ast} >  3\times 10^{10}$M$_{\odot}$ in our field as a function of  redshift, and the infrared luminosity function of Magnelli et al. (2009) for the number of LIRGs in the field. 
Given the relation between SFR and stellar mass at $z\sim 1$ calibrated by Elbaz et al. (2007), we expect the average SFR of a M$^\ast=3\times 10^{10} $M$_{\odot}$  galaxy will be $20\pm 10 $M$_{\odot}$/yr. 
With our LIRG selection, equivalent to $SFR > 18 $M$_{\odot}$/yr we expect that most LIRGs lie above our mass threshold, but not all of them, given the scatter of the relation.  
Therefore we can compute from the luminosity and mass functions (LF and MF)  an upper limit to our observed relation. The solid curve in Figure~\ref{frmipsz} shows the expected upper limit variation for the LIRG fraction. We find a  similar behavior, with the LIRG fraction increasing with redshift, but at a lower rate than what the LF and MF would predict: between $z=0.5$ and $z=1$, we observe a 3 fold increase in the LIRG fraction, while the computed increase is 4 fold. This discrepancy is expected since the solid line is an upper limit.
We plot with open circles the evolution of the LIRG fraction in low density environment ($5 < \Sigma_{10} < 10$ Mpc$^{-2}$), and with solid circles that for high density environment ($\Sigma_{10} > 90$ Mpc$^{-2}$). We find that the LIRG fraction evolves faster (a factor of 2) in high density environments than in low density ones, at 2$\sigma$ confidence level.  However, given the large errors, the high density environment fractions are still consistent with the low density environment fractions within 3 $\sigma$.

\section{CONCLUSIONS}

We studied the properties of a sample of LIRGs detected at 24 \micron\ in the COSMOS survey and with stellar masses M$^\ast>3\times 10^{10} $M$_{\odot}$ and $z=0.3\div 1.2$, as a function of the environment, by using accurate photometric redshifts.
In order to disentangle environment from any other possible effect  (mainly mass segregation) we studied the properties of our infrared galaxy sample in two bins of stellar mass.
The environment was characterized by using the distance of the 10th nearest neighbor, and the derived projected galaxy density  around each galaxy in redshift slices.  
Our results can be summarized as follows:
\begin{itemize}
\item The median SFR and SSFR per LIRG increase with redshift, in agreement with the expected evolution of the LIRG luminosity and mass functions (Le Floc'h et al. 2005, Caputi et al. 2007). 
The median SFR and median SSFR are constant at all densities within the errors, in agreement with what found by Hwang et al. (2010) in local LIRGs. 
We do not observe the reversal of the SFR density relation up to $z\sim 1$ for our mass--selected LIRG sample. 
Such reversal, if any, might occur at $z>1$ for high--mass LIRGs. 
This is not necessarily true for galaxies of smaller stellar mass. 

\item We find that the fraction of massive LIRGs among optically selected galaxies with the same stellar mass, varies as a function of both redshift and environment. 
The global LIRG fraction increases with increasing redshift, as expected from the cosmic evolution of the SFR density and of the infrared LF.  
The LIRG fraction decreases with increasing environment density.  Such decrease flattens going from $z=0.4$ to $z\sim 1$ and the fraction is constant in all environments for $z>1$. These results imply that the evolution of the LIRG fraction among optically selected galaxies from $z=1$ to $z=0.4$ is roughly twice as fast in dense than in low density environments.

\item Finally, we find that a large fraction of LIRGs is present in the most dense environments by $z\sim 1$. These might be the progenitors of the  massive galaxies found today in the center of clusters, which therefore were not likely formed through dry mergers.
\end{itemize}

Using photometric redshift, with their inherent limitations, we have shown that LSS and environment has a role on star formation. Future works based on spectroscopic surveys will refine these results in the densest structures, and open the realm of low density where our estimator is limited.

\acknowledgments
This work is based on observations made with the \emph{Spitzer Space Observatory}, which is operated by the Jet Propulsion Laboratory (JPL) , California Institute of Technology  under NASA contract 1407. 
Support for this work was provided by ANR "D--SIGALE" ANR--06--BLAN--0170 grant.
Support for this work was provided in part by NASA through contracts 1282612 and 1298231 issued by the JPL, and grants 1289085, 1310136.
We thank the anonymous referee for carefully reading and providing useful suggestions that improved the paper.

\clearpage

\end{document}